\documentclass{aa}
\usepackage{txfonts}
\usepackage{graphicx}
\usepackage{color,enumerate}

\usepackage[normalem]{ulem}

\usepackage{natbib}
\bibpunct{(}{)}{;}{a}{}{,} 

\def\be{\begin{equation}}
\def\ee{\end{equation}}
\def\bea{\begin{eqnarray}}
\def\eea{\end{eqnarray}}
\def\d{\mbox{d}}

\def\p{\partial}

\def\pder#1#2{\frac{\partial #1}{\partial #2}}

\def\vphi{{\cal V}^{(\phi)}}

\let\phi=\varphi
\let\rho=\varrho

\begin{document}

\title{LNRF-velocity hump-induced oscillations of a~Keplerian disc orbiting
  near-extreme Kerr black hole: A~possible explanation of high-frequency
  QPOs in GRS~1915+105}

\titlerunning{LNRF-velocity hump induced oscillations in GRS~1915+105}

\author{Zden\v{e}k Stuchl\'{\i}k \and Petr Slan\'{y} \and Gabriel
  T\"{o}r\"{o}k} 
\institute{Institute of Physics, Faculty of Philosophy and Science, Silesian
  University in Opava, Bezru\v{c}ovo n\'{a}m. 13,
  \\
  CZ-74601 Opava, Czech Republic}

\offprints{Z. Stuchl\'{\i}k, \\ \email{zdenek.stuchlik@fpf.slu.cz}}

\date{Received / Accepted}
\keywords{Accretion, accretion discs -- Black hole physics -- Methods: data
  analysis}

\abstract
{At least four high-frequency quasiperiodic oscillations (QPOs) at frequencies
41\,Hz, 67\,Hz, 113\,Hz, and 167\,Hz were reported in a binary system GRS~1915+105
hosting near-extreme Kerr black hole with a dimensionless spin $a>0.98$.
}
{We attempt to explain all four observed frequencies by an extension of the
 standard resonant model of epicyclic oscillations.
} 
{We use the idea of oscillations induced by the hump of the orbital velocity
profile (related to locally non-rotating frames--LNRF) in discs orbiting near-extreme Kerr black holes, which are characterized by a~``humpy frequency'' $\nu_{\rm h}$, that could excite the radial and vertical epicyclic oscillations with frequencies $\nu_{\rm r}$, $\nu_{\rm v}$. Due to non-linear resonant phenomena, the combinational frequencies are allowed as well.
}
{Assuming mass $M=14.8\,M_{\odot}$ and spin $a=0.9998$ for the GRS~1915+105
Kerr black hole, the model predicts frequencies $\nu_{\rm h}=41$\,Hz, $\nu_{\rm
  r}=67$\,Hz, $\nu_{\rm h}+\nu_{\rm r}=108$\,Hz, and $\nu_{\rm v}-\nu_{\rm
  r}=170$\,Hz corresponding quite well to the observed ones.
} 
{For black-hole parameters being in good agreement with those given
observationally, the forced resonant phenomena in non-linear oscillations, 
excited by the "hump-induced" oscillations in a Keplerian disc, can explain
high-frequency QPOs in near-extreme Kerr black-hole binary system 
GRS~1915+105 within the range of observational errors.
}

\maketitle 


\section{Introduction}
Detailed analysis of the variable X-ray black-hole binary system (microquasar)
GRS~1915+105 reveals high-frequency QPOs appearing at four frequencies, namely
$\nu_{1}=(41\pm 1)$\,Hz, $\nu_{2}=(67\pm 1)$\,Hz
\citep{Mor-Rem-Gre:1997:ASTRJ2:,Str:2001:ASTRJ2:}, and $\nu_{3}=(113\pm
5)$\,Hz, $\nu_{4}=(167\pm 5)$\,Hz \citep{Rem:2004:AIPC:}. In this range of its
errors, both pairs are close to the frequency ratio 3:2 suggesting
the possible existence of resonant phenomena in the system. Observations of
oscillations with these frequencies have different qualities, but in all four
cases the data are quite convincing; see 
\citep{McCli-Rem:2004:CompactX-Sources:,Rem-McCli:2006:ARASTRA:}. 

Several models have been developed to explain the kHz QPO frequencies, and it is usually preferred that these oscillations are related to the orbital motion near the inner edge of an accretion disc. In particular, two ideas based on the strong-gravity properties have been proposed. While \citet{Ste-Vie:1998:ASTRJ2:,Ste-Vie:1999:PHYRL:} introduced the ``Relativistic Precession Model'' considering that the kHz QPOs directly manifest the modes of a slightly perturbed (and therefore epicyclic) relativistic motion of blobs in the inner parts of the accretion disc, \citet{Klu-Abr:2001:ACTPB:} propose models based on non-linear oscillations of an accretion disc that assume resonant interaction between orbital and/or epicyclic modes. In a different context, the possibility of resonant coupling between the epicyclic modes of motion in the Kerr spacetime was also mentioned in the early work of \citet{Ali-Gal:1981:GENRG2:}.

In the case of near-extreme Kerr black holes, it was suggested that the
epicyclic oscillations in the disc could be excited by resonances with the
so-called ``hump-induced'' oscillations, see papers of 
\citet{Asch:2004:ASTRA:,Asch:2006:CHIJAA:} and 
\citet{Stu-Sla-Tor:2004:RAGtime4and5:,Stu-Sla-Tor:2007a:ASTRA:}. 
This idea was proposed so as to extend standard orbital (resonant) models meant to
explain high-frequency QPOs observed in black-hole sources. 

Recently, careful and detailed analysis of the spectral continuum from
GRS~1915+105 has put a strong limit on the black-hole spin,\footnote{Units
  $c=G=M=1$ ($M$ is the total mass of the Kerr black hole) and the
  Boyer-Lindquist (B-L) coordinates $(t,\,r,\,\theta,\,\phi)$ are used
  hereafter.} 
namely $0.98 < a < 1$ \citep{McCli-etal:2006:ASTRJ2:}, indicating the
presence of near-extreme Kerr black hole whose mass has been restricted
observationally to $M=(14.0\pm 4.4)\, M_{\odot}$, see
\citep{McCli-Rem:2004:CompactX-Sources:,Rem-McCli:2006:ARASTRA:}. 
Therefore, the microquasar GRS~1915+105 seems to be an appropriate candidate
to test the extended resonant model with hump-induced
oscillations.\footnote{However, \citet{Mid-etal:2006:MONNR:} refer to a 
  substantially lower, intermediate value of black-hole spin, $a\sim 0.7$, to
  which the model of hump-induced oscillations cannot be applied.} 

The idea of hump-induced oscillations and their possible resonant coupling with
the epicyclic ones is briefly discussed in Sect.~\ref{s2}. The related resonant
model, assuming the excitation of epicyclic oscillations by the hump-induced
oscillations through non-linear resonant phenomena, is applied to GRS~1915+105
in Sect.~\ref{s3}, concluding remarks are presented in Sect.~\ref{s4}.

\section{Hump-induced and epicyclic oscillations in Keplerian discs and 
  possible resonant coupling}\label{s2}
In order to describe the local processes in an accretion disc, it is necessary to choose a local observer (characterized by its reference frame). In general relativity there is no preferred observer. On the other hand, if we want to study processes related to the orbital motion of matter in the disc, it is reasonable to choose the observers with zero angular momentum, so-called ZAMOs, as their reference frames do not rotate with respect to the spacetime, and thus ZAMOs should reveal local kinematic properties of the disc in the clearest way. (In rotating--stationary, axisymetric--spacetimes, they are dragged along with the spacetime.) In the Kerr spacetime, ZAMOs are represented by locally non-rotating frames (LNRF); see \citet{Bar-Pre-Teu:1972:ASTRJ2:}. Notice that in the Schwarzschild spacetime, LNRF correspond to the static observer frames. 

\cite{Asch:2004:ASTRA:} finds that for near-extreme Kerr black holes with
the spin $a > 0.9953$, the test-particle orbital velocity ${\cal V}^{(\phi)}$ related to LNRF reveals a hump in the equatorial plane ($\theta=\pi/2$). This non-monotonicity is located in a small region inside the ergosphere of the black-hole spacetime close to, but above, the marginally stable orbit.\footnote{
We stress that the Aschenbach effect is frame-dependent, as it is related to LNRF, but recall the arguments for relevance of the LNRF point of view at the beginning of the section.
}
Therefore, it can be relevant for thin accretion discs around near-extreme Kerr black holes, as the inner edge of the disc can extend down to the innermost stable circular orbit (ISCO). 

Moreover, \cite{Stu-Sla-Tor-Abr:2005:PHYSR4:} shows that for $a>0.99979$ the
similarly humpy behavior of the orbital velocity in LNRF also takes place for the non-geodesic motion of test perfect fluid in marginally stable barotropic tori characterized by the uniform
distribution of the specific angular momentum,
$\ell(r,\,\theta)\equiv -U_{\phi}/U_{t}=\mbox{const}$, where the motion of
fluid elements is given by the 4-velocity field 
$U^{\mu}=(U^{t}(r,\,\theta),\,0,\,0,\,U^{\phi}(r,\,\theta))$. Outside the
equatorial plane, the non-monotonic behavior of ${\cal V}^{(\phi)}$ in
marginally stable tori is represented by the topology change of the cylindrical
equivelocity surfaces in the region of the hump, because the toroidal equivelocity surfaces centered around the circle corresponding to the local minimum of ${\cal V}^{(\phi)}$ in the equatorial plane exist for $a>0.99979$ \citep{Stu-Sla-Tor-Abr:2005:PHYSR4:}. This suggests
a~generation of possible instabilities in radial and vertical directions; see
\citet{Stu-Sla-Tor:2004:RAGtime4and5:}. In the following, we restrict our attention to the case of Keplerian discs. 

\begin{figure}
\centering
\includegraphics[width=1 \hsize]{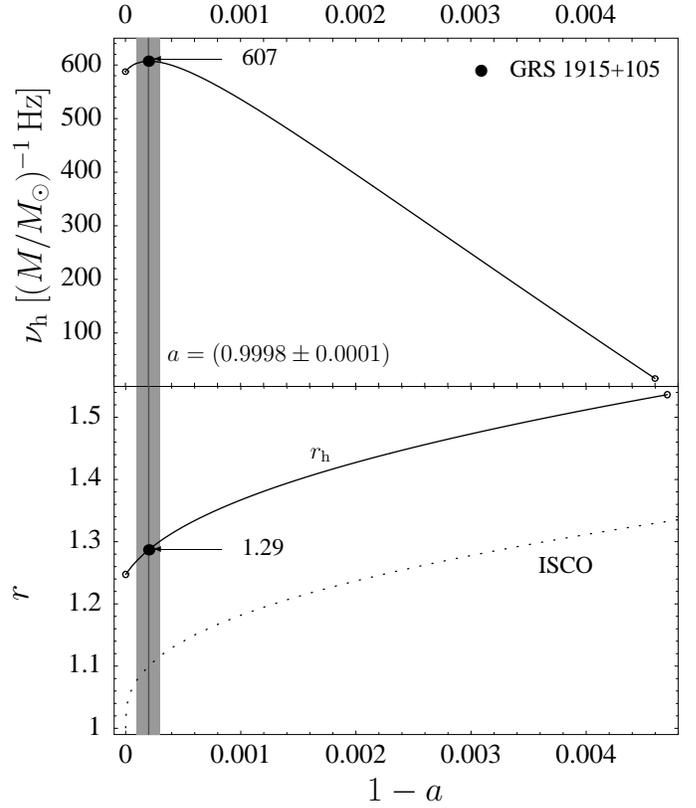}
\caption{Spin-dependence of the humpy frequency $\nu_{\rm h}$ and the humpy
  radius $r_{\rm h}$. For completeness, the B-L radius of the innermost stable
  circular geodesic (ISCO) is plotted.}
\label{f1}
\end{figure}

Heuristic connection between the positive part of the velocity gradient,
$\p{\cal V}^{(\phi)}/\p r$, and the excitation of epicyclic oscillations in
Keplerian discs was suggested by \citet{Asch:2004:ASTRA:,Asch:2006:CHIJAA:},
who defined the characteristic frequency of oscillations, induced by the humpy
profile of ${\cal V}^{(\phi)}$, by the maximum positive
slope of the orbital velocity in terms of the coordinate radius, $\nu_{\rm
  crit}\equiv(\p{\cal V}^{(\phi)}/\p r)_{\rm max}$. This coordinate-dependent
definition was corrected in \citet{Stu-Sla-Tor:2007a:ASTRA:}, where the proper
radial distance $\d\tilde{r}=\sqrt{g_{rr}}\,\d r$ rather than the coordinate
distance $\d r$ was used to define the characteristic (critical) frequency 
$\nu_{\rm crit}^{\tilde{r}}\equiv(\p{\cal V}^{(\phi)}/\p\tilde{r})_{\rm max}$.
Such a locally defined critical frequency was further related to a stationary
observer at infinity, obtaining the so-called ``humpy frequency''
\bea                                                               \label{e1}
     \nu_{\rm h} &=& \sqrt{-(g_{tt}+2\omega
       g_{t\phi}+\omega^2 g_{\phi\phi})_{r=r_{\rm h}}} \,\nu^{\tilde{r}}_{\rm
       crit}  
                                                                  \nonumber \\
     &=& \sqrt{\frac{1}{r_{\rm h}}\left
       [(r_{\rm h}-2)-\frac{4a^2}{r_{\rm h}(r_{\rm h}^{2}+a^2)+2a^2}\right ]} 
                                                                  \nonumber \\
     &\times& \left [-\frac{r_{\rm h}^{5}+a^4(3r_{\rm h}+2)-2a^3
      r_{\rm h}^{1/2}(3r_{\rm h}+1)}{2\Delta_{\rm h}^{3/2}\sqrt{r_{\rm
          h}}(r_{\rm h}^{3/2}+a)^2} \right. \nonumber \\ 
    &-& \left. \frac{2a^2 r_{\rm h}^{2} (2r_{\rm h}-5)-2ar_{\rm
          h}^{5/2}(5r_{\rm h}-9)}{2\Delta_{\rm h}^{3/2}\sqrt{r_{\rm h}} 
        (r_{\rm h}^{3/2}+a)^2}\right ]\frac{\sqrt{\Delta_{\rm h}}}{r_{\rm h}}, 
\eea
where $g_{\mu\nu}$ are the metric coefficients of the Kerr geometry and
$\omega=-g_{t\phi}/g_{\phi\phi}$ is the angular velocity of the LNRF; see,
e.g., \citet{Bar-Pre-Teu:1972:ASTRJ2:}; $\Delta_{\rm h} = r_{\rm h}^2-2r_{\rm
  h}+a^2$. The analytic formula is given for the equatorial plane
($\theta=\pi/2$). The B-L radius $r_{\rm h}$ where the positive gradient of
the velocity profile in terms of the proper radial distance reaches its
maximum, so-called ``humpy radius'', is given by the condition
\be                                                                \label{e2}
\pder{}{r}\left( \pder{\vphi}{\tilde{r}}\right )=0
\ee
leading to the equation
\bea                                                               \label{e4}
\lefteqn{3a^7(r+2)+a^6\sqrt{r}(21r^2+18r-4)-a^5 r(33r^2+10r+20)} \nonumber\\
\lefteqn{+a^4 r\sqrt{r}(45r^3-62r^2-68r+16)-a^3 r^3(83r^2-122r-60)} \nonumber\\
\lefteqn{+a^2 r^4 \sqrt{r}(27r^2-130r+136)-9a r^5(7r^2-26r+24)} \nonumber\\
\lefteqn{+r^7 \sqrt{r}(3r-2)=0,}
\eea
which must be solved numerically. The spin dependence of the humpy radius and the
related humpy frequency is illustrated in Fig.~\ref{f1}.
The humpy radius $r_{\rm h}$ falls monotonically with increasing spin $a$,
while the humpy frequency $\nu_{\rm h}$ has a maximum for $a=0.9998$, where
$\nu_{\rm h\,(max)}=607\,(M_{\sun}/M)$\,Hz, and it tends to $\nu_{{\rm
    h}\,({a\,\rightarrow1})}=588\,(M_{\sun}/M)$\,Hz.

When particles following a Keplerian circular orbit are perturbed, they begin
to follow, in the first approximation, an epicyclic motion around the
equilibrium Keplerian orbit, generally characterized by the frequencies of the 
radial and vertical epicyclic oscillations $\nu_{\rm{r}}, \nu_{\rm{v}}$
\citep{Ali-Gal:1981:GENRG2:,Now-Leh:1998:TheoryBlackHoleAccretionDisks:}:
\bea                                                               \label{e5}
     \nu^2_{\rm r} &=& \nu^2_{\rm K}(1-6r^{-1} + 8ar^{-3/2} - 3a^2r^{-2}),
        \quad \\                                                   \label{e6}
        \nu^2_{\rm v} & \equiv & \nu^2_{\theta} = \nu^2_{\rm K} (1-
        4ar^{-3/2} + 3a^2r^{-2}),       
\eea
where $\nu_{\rm K}$ is the Keplerian orbital
frequency
\be                                                                \label{e7}
\nu_{\rm K} = \frac{1}{2\pi(r^{3/2}+a)}.
\ee
The ratios of the humpy frequency and the epicyclic frequencies at
the humpy radius were determined in \citet{Stu-Sla-Tor:2007a:ASTRA:}
revealing almost spin-independent asymptotic behavior for $a\to 1$
represented closely by the ratios of integer numbers, $\nu_{\rm 
  v}:\nu_{\rm r}:\nu_{\rm h}\sim 11:3:2$, which imply a possibility of
resonant phenomena between the hump-induced and epicyclic oscillations
predicted by \citet{Asch:2004:ASTRA:}. The ratios of the epicyclic 
frequencies and the humpy frequency are given in the dependence on the 
black-hole spin in Fig.~\ref{f2}.

\section{Application of the hump-induced resonance model to high-frequency
  QPOs in GRS~1915+105}\label{s3}
\begin{figure}
\centering
\includegraphics[width=1 \hsize]{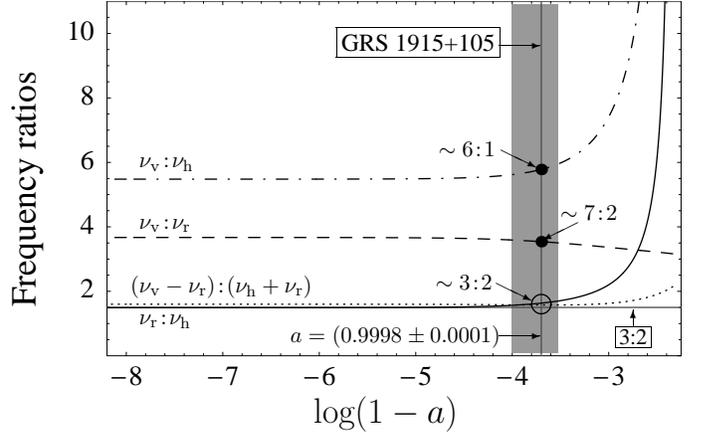}
\caption{Spin dependence of frequency ratios including the radial ($\nu_{\rm
    r}$) and vertical ($\nu_{\rm v}$) epicyclic frequencies, and the humpy
    frequency ($\nu_{\rm h}$) evaluated at the same radius $r_{\rm h}$ where
    the humpy frequency is defined. The range of the spin relevant for
    GRS~1915+105 Kerr black hole is shaded. For the mean value $a=0.9998$, the
    frequency ratios are close to the ratios of integer numbers, suggesting a
    possibility of resonances between hump-induced and epicyclic oscillations
    in GRS~1915+105.}
\label{f2}
\end{figure}

Primarily concentrating on the lower pair of frequencies, we assume that the
lowest frequency is directly the humpy frequency,
\be                                                                \label{e8}
\nu_{\rm h}\equiv \nu_1 = (41 \pm 1)\,\mbox{Hz},
\ee
while the second lowest frequency corresponds directly to the radial epicyclic
frequency at the same radius $r_{\rm h}$,
\be                                                                \label{e9}
\nu_{\rm r}\equiv \nu_2 = (67\pm 1) \,\mbox{Hz}.
\ee
These frequencies are close to a 3:2 ratio, therefore the forced
non-linear resonance can be relevant in such a situation. 
The ratio of $\nu_{\rm r}/\nu_{\rm h} = (1.63\pm 0.06)$ gives the black hole
spin $a=(0.9998 \pm 0.0001)$ (the uncertainty of the spin is implied by
uncertainties of the lower pair of frequencies being $\sim 1$\,Hz); see
Fig. \ref{f2}. Notice that this spin corresponds to the maximal possible value of the
humpy frequency $\nu_{\rm h\,(max)}$ (Fig. \ref{f1}). Since the humpy
frequency is $1/M$-scaled, the absolute value of $\nu_{\rm h}$ implies the
black hole mass $M=(14.8\pm 0.4)\, M_{\sun}$. The corresponding humpy radius is
$r_{\rm h}= 1.29\,^{+0.01}_{-0.02}$ (Fig.~\ref{f1}).
At such a~radius, the vertical epicyclic frequency of a~particle orbiting the
Kerr black hole with the mass and spin inferred above reaches the value $\nu_{\rm
  v}=(0.23\pm 0.01)$\,kHz. Then the upper pair of observed frequencies can be
explained, within the range of observational errors $\pm 5$\,Hz, by
combinational frequencies at the humpy radius $r_{\rm h}$ in the following way:
\bea
\nu_{\rm 3}\sim(\nu_{\rm r}+\nu_{\rm h}) &=& (108\pm 2)\,\mbox{Hz} 
                                                                 \label{e10} \\
\nu_{\rm 4}\sim(\nu_{\rm v}-\nu_{\rm r}) &=& (0.17\pm 0.01)\,\mbox{kHz}. 
                                                                 \label{e11}
\eea 

\section{Conclusions}\label{s4}
The idea of epicyclic oscillations induced by the LNRF-velocity hump in the
region where the positive part of the velocity gradient reaches its maximum is
able to address all four high-frequency QPOs observed in the X-ray source
GRS~1915+105.
The model implies a near-extreme spin of the central black hole ($a\sim
0.9998$), which agrees well with results from the spectral
continuum fits, and the black-hole mass $M\sim 14.8\,M_{\sun}$ being
well inside the interval given by other observational methods.
Note that the orbital resonance model of Klu\'{z}niak \& Abramowicz, assuming the parametric resonance between the vertical and radial epicyclic oscillations in frequency ratio 3:2 represented by the upper pair of observed frequencies, also gives the spin $a>0.99$ but for $M\simeq 18M_{\odot}$ \citep{Tor-Abr-Klu-Stu:2005:ASTRA:}. On the other hand, the ``Relativistic Precession Model'' gives a~substantially lower value for the spin: $a\sim 0.3$ \citep{Ste-Vie-Mor:1999:ASTRJ2:}.

In the presented model, we assume that all four observed frequencies arise due
to forced non-linear oscillations of the Keplerian disc at the same radius
$r_{\rm h}$, excited by the hump-induced oscillations characterized by the
humpy frequency $\nu_{\rm h}$.
The black-hole parameters $a,\,M$ are fixed by the requirement that
the lower pair of observed frequencies is identified with the humpy
frequency and the radial epicyclic frequency, $\nu_1 \equiv \nu_{\rm h}$,
$\nu_2 \equiv \nu_{\rm r}$. Assuming non-linear resonant phenomena enabling
the existence of combinational frequencies and the possibility of observing them,
the upper pair of observed frequencies can be explained as the combinational
ones of the humpy frequency and both epicyclic frequencies,
$\nu_{3}\sim(\nu_{\rm r}+\nu_{\rm h})$, $\nu_{4}\sim(\nu_{\rm v}-\nu_{\rm r})$.
Moreover, both frequency ratios $\nu_{\rm r}:\nu_{\rm h}$,
and $(\nu_{\rm v}-\nu_{\rm r}):(\nu_{\rm r}+\nu_{\rm h})$ are close to 3:2
ratio (Fig.~\ref{f2}), in which the resonant phenomena can be strong
enough. On the other hand, as $4\nu_{\rm h}=(164\pm 4)$\,Hz, which is also
close to the uppermost frequency, there is another possibility of explaining
$\nu_4$ through a~sub-harmonic resonance forced by the humpy oscillations as
well. Finally, note that \citet{Str:2001:ASTRJ2:} also reports another
relatively weak QPO at frequency of $(56\pm 2)$\,Hz. If this is the case
(which, according to our knowledge, has not been confirmed by other
observations yet), it could be related to the second harmonic of the
combinational frequency\footnote{Combinational frequency $(\nu_{\rm
    r}-\nu_{\rm h})$ corresponds to the same order of nonlinearity as
  $(\nu_{\rm r}+\nu_{\rm h})$. \\
\emph{Note added in the manuscript:} After the paper was accepted we obtained an information that a weak QPO at frequency 27\,Hz is referenced in \citet{Bel-Men-San:2001:ASTRA:}.} $(\nu_{\rm r}-\nu_{\rm h})=(26\pm 2)$\,Hz.

Generally, other harmonics and combinational frequencies may occur in
a~non-linear oscillating system corresponding to higher approximations, when
the equation of motion describing the non-linear oscillations is solved by the
method of successive approximations. 
The statement by \citet{Lan-Lif:1976:Mechanics:} that 
``As the degree of approximation increases, however, the strength of the resonances, and the widths of the frequency ranges in which they occur, decrease so rapidly that in practice
only the resonances at frequencies\footnote{$p,\,q$ are integers.}
$\nu\approx p\nu_{0}/q$ with small $p$ and $q$ can be observed'' can explain why
a~QPO near the frequency 237~Hz, corresponding to the
vertical epicyclic frequency $\nu_{\rm v}$ at the same radius 
$r_{\rm h}$ as the previously mentioned humpy and radial epicyclic frequencies
$\nu_{\rm h},\,\nu_{\rm r}$, is not directly observed, despite the
commensurability of these frequencies represented by the frequency ratios
$\nu_{\rm v}:\nu_{\rm h}\sim 6:1$ and $\nu_{\rm v}:\nu_{\rm r}\sim 7:2$
(Fig.~\ref{f2}).

\begin{acknowledgements}
The authors are supported by the Czech grant MSM~4781305903.
\end{acknowledgements}

\bibliographystyle{aa} 


\end{document}